# PROJECT X FUNCTIONAL REQUIREMENTS SPECIFICATION*

S. D. Holmes, S. D. Henderson, R. Kephart, J. Kerby, S. Mishra, S. Nagaitsev, R. Tschirhart
Fermilab, Batavia, IL, 60510, U.S.A.

*Abstract*

Project X is a multi-megawatt proton facility being developed to support intensity frontier research in elementary particle physics, with possible applications to nuclear physics and nuclear energy research, at Fermilab. A Functional Requirements Specification has been developed in order to establish performance criteria for the Project X complex in support of these multiple missions. This paper will describe the Functional Requirements for the Project X facility and the rationale for these requirements.

## PROJECT X MISSION AND GOALS

Project X is a high intensity proton facility being developed to support a world-leading U.S. program in Intensity Frontier over the next several decades. Project X is an integral part of the long range strategic plan for the U.S. Department of Energy (DOE) High Energy Physics program as described in the P5 report of May 2008 [1], and the Fermilab Steering Group Report of August 2007 [2].

The primary mission elements to be supported by Project X include:

1. Provdee a neutrino beam for long baseline neutrino oscillation experiments, based on targeting at least 2 MW of proton beam power at any energy between 60 – 120 GeV.
2. Provide MW-class, multi-GeV, proton beams supporting multiple kaon, muon, and neutrino based precision experiments. Simultaneous operations of the rare processes and neutrino programs are required.
3. Provide a path toward a muon source for a possible future Neutrino Factory and/or a Muon Collider.
4. Provide options for implementing a program of Standard Model tests with nuclei and/or nuclear energy applications

These elements represent the fundamental design criteria for Project X.

The development of a design concept for a high intensity proton facility has gone through several iterations culminating in a concept, designated the Project X Reference Design [3], that meets the high level design criteria listed above in an innovative and flexible manner. The Reference Design is based on a 3 GeV superconducting CW linac, augmented by a superconducting pulsed linac for acceleration from 3-8 GeV, and modifications to the existing Recycler and Main Injector Rings at Fermilab. The Reference Design provides a facility that will be unique in the world with unmatched capabilities for the delivery of very high beam power with flexible beam formats to multiple users.

## PROJECT X REFERENCE DESIGN

Figure 1 presents a schematic depiction of the Project X Reference Design aligned with the mission elements outlined in Section I. The primary elements are:

- An H- source consisting of a 2.5 MeV RFQ, and Medium Energy Beam Transport (MEBT) augmented with a wideband chopper capable of accepting or rejecting bunches in arbitrary patterns at up to 162.5 MHz;
- A 3 GeV superconducting linac operating in CW mode, and capable of accelerating an average (averaged over >1 μsec) beam current of 1mA, and a peak beam current (averaged over <1 μsec) of 5 mA;
- A 3 to 8 GeV pulsed superconducting linac capable of accelerating an average current of 33 μA with a 1-5% duty cycle;
- A pulsed dipole that can split the 3 GeV beam between the neutrino program and the rare processes program;
- An rf beam splitter that can deliver the 3 GeV beam to multiple (at least three) experimental areas;
- Experimental facilities and detectors to support an initial round of 3 GeV experiments;
- Modification to the Recycler and Main Injector Ring required to support delivery of 2 MW of beam power from the Main Injector at any energy between 60-120 GeV;
- All interconnecting beam transport lines.

## PROJECT X FUNCTIONAL REQUIREMENTS

In order to guide the development of the Reference Design into a complete Conceptual Design a set of Functional Requirements have been established for Project X. These requirements are based on the mission elements described above and are developed within the context of the Reference Design displayed in Figure 1. In addition the Functional Requirements are subject to certain assumptions, interface requirements, and constraints. Principal among these are:

---

* Work supported by the Fermi Research Alliance, under contract to the U.S. Department of Energy

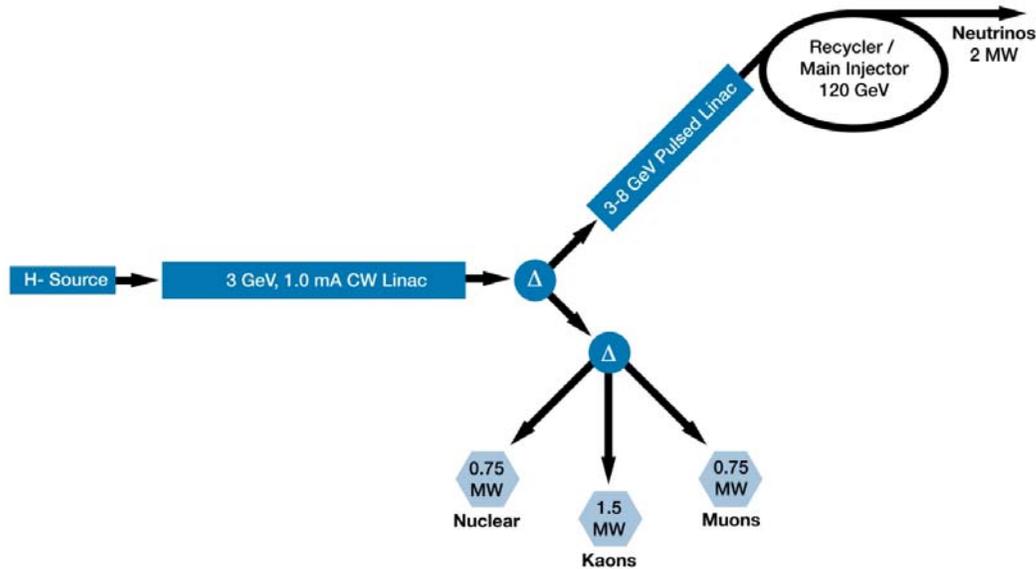

Figure 1: Project X Reference Design

- Project X will be constructed on the Fermilab site.
- The long baseline neutrino program will eventually require beam power upgrades beyond the 2 MW initial capability..
- The rare process program requires MW class, high duty factor, proton beams with flexible bunch patterns. This is best accomplished with a CW linac based on superconducting rf technology.
- Production of adequate number of kaons sets the minimum CW linac energy at ~ 3 GeV
- The neutrino and 3 GeV programs must operate simultaneously.
- The Recycler Ring operates at a fixed energy of 8 GeV; the upgraded Main Injector will be capable of injection energies as low as 6 GeV.
- The NOvA Project will have upgraded the Main Injector /Reccyler to support 700 kW operations.
- A pulsed H- linac for acceleration from 3 to 8 GeV provides a flexible base for future development of the Fermilba accelerator complex.
- The 3 GeV CW linac must be capable of accelerating $H^-$ ions.
- Project X construction and commissioning will be a national project with international participation. with Fermilab as the lead laboratory.
- Project X technology choices will exploit the world-wide ILC and Muon Collider R&D activity, provided such choices do not significantly degrade the technical performance, cost, or upgrade path.
- Project X represents a large long-term investment for U.S. HEP and as such must be robust, flexible, and designed with significant upgrade potential.

The Functional Requirements for Project are summarized in Table 1. There are five groups of requirements corresponding to: 1)the 3 GeV CW linac; 2)the 3-8 GeV pulsed linac; 3)the Recycler/Main Injector Complex; 4)integration requirements; and 5)upgradability requirements.

## SUMMARY

Project X is central to Fermilab's strategy for future development of the accelerator complex. Project X will support a world leading program in neutrinos and other rare processes over the coming decades, and will be constructed in a manner that could provide a stepping stone to muon-based facilities – either a Neutrino Factory or a Muon Collider. A reference design has been completed which satisfies the four primary mission elements and provides a flexible platform for future development of the Fermilab accelerator complex. Based on the reference design a Functional Requirements Specification has been developed that will form the basis of completion of a complete Conceptual Design.

Table 1: Project X Functional Requirements

| Requirement | Description | Value |
|---|---|---|
| **3 GeV Linac** | | |
| L1 | Delivered Beam Energy, maximum | 3 GeV (kinetic) |
| L2 | Delivered Beam Power at 3 GeV | 3 MW |
| L3 | Average Beam Current (averaged over >1 μsec) | 1 mA |
| L4 | Maximum Beam Current (sustained for <1 μsec) | 5 mA |
| L5 | The 3 GeV linac must be capable of delivering correctly formatted beam to the pulsed linac | |
| L6 | Charge delivered to pulsed linac | 26 mA-msec in < 0.75 sec |
| L7 | Maximum Bunch Intensity | $1.9 \times 10^8$ |
| L8 | Minimum Bunch Spacing | 6.2 nsec (1/162.5 MHz) |
| L9 | Bunch Length | <50 psec (full-width half max) |
| L10 | Bunch Pattern | Programmable |
| L11 | RF Duty Factor | 100% (CW) |
| L12 | RF Frequency | 162.5 MHz and harmonics thereof |
| L13 | 3 GeV Beam Split | Three-way |
| **3-8 GeV Pulsed Linac** | | |
| P1 | Maximum Beam Energy | 8 GeV |
| P2 | The 3-8 GeV pulsed linac must be capable of delivering correctly formatted beam for injection into the Recycler Ring (or Main Injector) | |
| P3 | Charge to fill Main Injector/cycle | 26 mA-msec in <0.75 sec |
| P4 | Maximum beam power delivered to 8 GeV | 300 kW |
| P5 | Duty Factor (initial) | < 4% |
| **Main Injector/Recycler** | | |
| M1 | Delivered Beam Energy, maximum | 120 GeV |
| M2 | Delivered Beam Energy, minimum | 60 GeV |
| M3 | Minimum Injection Energy | 6 GeV |
| M4 | Beam Power (60-120 GeV) | > 2 MW |
| M5 | Beam Particles | Protons |
| M6 | Beam Intensity | $1.6 \times 10^{14}$ protons per pulse |
| M7 | Beam Pulse Length | ~10 μsec |
| M8 | Bunches per Pulse | ~550 |
| M9 | Bunch Spacing | 18.8 nsec (1/53.1 MHz) |
| M10 | Bunch Length | <2 nsec (fullwidth half max) |
| M11 | Pulse Repetition Rate (120 GeV) | 1.2 sec |
| M12 | Pulse Repetition Rate (60 GeV) | 0.75 sec |
| M13 | Max Momentum Spread at extraction | $2 \times 10^{-3}$ |
| **Integration** | | |
| I1 | The 3 GeV and neutrino programs must operate simultaneously | |
| I2 | Residual Activation from Uncontrolled Beam Loss in areas requiring hands on maintenance. | <20 mrem/hour (average) <100 mrem/hour (peak) @ 1 ft |
| I3 | Scheduled Maintenance Weeks/Year | 8 |
| I4 | 3 GeV Linac Operational Reliability | 90% |
| I5 | 60-120 GeV Operational Reliability | 85% |
| I6 | Facility Lifetime | 40 years |
| **Upgradability** | | |
| U1 | Provisions should be made to support an upgrade of the CW linac to an average current of 4 mA. | |
| U2 | Provisions should be made to support an upgrade of the Main Injector to a delivered beam power of ~4 MW at 120 GeV. | |
| U3 | Provisions should be made to deliver CW proton beams as low as 1 GeV. | |
| U4 | Provision should be made to support an upgrade to the CW linac such that it can accelerate Protons. | |
| U5 | Provisions should be made to support an upgrade of the pulsed linac to a duty factor of 10%. | |
| U6 | Provisions should be made to support an upgrade of the CW linac to 3.1 ns bunch spacing. | |